\documentclass[%
 reprint,
 amsmath,amssymb,
 aps,
]{revtex4-1}

\usepackage{graphicx}
\usepackage{dcolumn}
\usepackage{bm}

\begin{document}

\preprint{APS/123-QED}

\title{Investigation of antineutrino spectral anomaly with updated nuclear database}
\thanks{This work was supported by National Natural Science Foundation of China (Grant No. 11390383) and the Fundamental Research Funds for the Central Universities(Grant No. 2018ZD10,2018MS044).}%

\author{Ma Xubo$^{1}$,}
\email{maxb@ncepu.edu.cn}
\author{Yang Le$^{1}$, Zhan Liang$^{2}$, An Fengpeng$^{3}$, Cao Jun$^{2}$ }
\affiliation{%
 $^{1} $School of Nuclear Science and Engineering, North China Electric Power University, Beijing,China. \\
 $^{2}$Institute of High Energy Physics, Beijing, China. \\
 $^{3}$Institute of Modern Physics, East China University of Science and Technology, Shanghai, China.
}%

\date{\today}

\begin{abstract}
Recently, three successful antineutrino experiments (Daya Bay, Double Chooz, and RENO) measured the neutrino mixing angle $\theta_{13}$; however, significant discrepancies were found, both in the absolute flux and spectral shape. In this study, the antineutrino spectra were calculated by using the updated nuclear structure library, and we found that the four isotopes antineutrino spectrum have all contribution to the 5--7 MeV bump with ENDF/B-VII.1 fission yield. The bump can be explained well using the updated library and more important isotopes contribution to the bump were also given. In the last, the fission yield correlation coefficient between the four isotopes were discussed, and found that the correlation coefficients are very large.

\end{abstract}

\pacs{Valid PACS appear here}
\maketitle


\section{Introduction}
Recently, three successful antineutrino experiments (Daya Bay\cite{Dayabay}, Double Chooz\cite{DoubleChooz}, and RENO\cite{RENO}) measured the neutrino mixing angle $\theta_{13}$; however, significant discrepancies were found, both in the absolute flux and spectral shape. The discrepancies of the absolute flux was called "reactor neutrino anomaly", which first appeared in a publication by Mention et al.\cite{mention}.  the antineutrino spectral shape, a 2.9$\sigma$ deviation was found in the measured inverse beta decay  position energy spectrum compared to predictions. In particular, an excess of events at energies of 4 -- 6 MeV was found in the measured spectrum\cite{spc1Dayabay,spc2Dayabay,spc2Chooz}, with a local significance of 4.4$\sigma$. These results have brought home the notion that neutrino fluxes are not as well understood as had been thought. At present, it is not clear what physical processes give rise to the neutrino spectra bump. Much effort has been focused on the reactor antineutrino anomaly, which arose from improved calculations of the antineutrino spectra derived from a combination of information from nuclear databases with reference $\beta$ spectra\cite{a8,a9,a10,a7,a11}.
In general, there are two approaches which were applied to estimate antineutrino spectral. One is called "$\beta^{-}$ conversion". The energy spectra of the $\beta^{-}$ from beta decay were measured to estimate the corresponding $\bar{\nu}_{e}$ emission for the fissile isotopes $^{235}$U,$^{239}$Pu,$^{241}$Pu from ILL reactor in the 1980s\cite{a8,a9,a10}. More recently, a similar measurement was made for $^{238}$U\cite{u238measure}. For a single measured $\beta^{-}$ decay spectrum, the corresponding $\bar{\nu}_{e}$ spectrum can be predicted with high precision. An other method was called "\emph{ab initial}  method". Using the measured $\beta^{-}$ decay parameters and fission yields of the nuclear library, the antineutrino spectra of each isotopes can be evaluated by summing each measured $\beta^{-}$ spectra of the fission daughters\cite{fallot}\cite{hayes}\cite{dwyer}. This introduces uncertainties of a few percent in the corresponding predictions of such calculation.

In this Letter, we discuss an alternative calculation of $\bar{\nu}_{e}$ spectrum from nuclear structure data library which combined the data from JENDL/DDF-2015 \cite{jendldecay}(1103 kinds of isotopes decay data) and JEFF3.1.1 nuclear database \cite{jeffdecay} (172 kinds of isotopes decay data which are not included in the JENDL library). The new database is coded updated nuclear database compared with JEFF3.1.1 data base. The fission yield library with different library, such as ENDF/B-VII.1, JEFF3.1.1 and JENDL4.0 are used, and the spectra ratio with different library are compared. The \emph{ab initio} approach was applied and 1275 isotopes nuclear data were used which cause large uncertainties. Despite these uncertainties, we find that \emph{ab initio} calculation predicting an excess $\bar{\nu}_{e}$'s with $E_{\bar{\nu}_{e}}$=5-7 MeV relative to the $\beta^{-} conversion $ method.
We also find that the bump is more notable with combined nuclear structure data library than that with only JENDL nuclear structure data library, and for the bump, the four isotopes all have contribution to the bump.

\section{Calculation of isotope antineutrino spectrum}
The \emph{ab initio} method of calculating the isotope antineutrino spectrum that presented in Refs.\cite{fallot,hayes,dwyer}. For a system in equilibrium, the total antineutrino spectra is given by
\begin{equation}
S(E_{\bar{\nu}})=\sum_{i=1}^{N}Y_{i}\sum_{j=1}^{M} f_{ij}S_{ij}(E_{\bar{\nu}})
\label{totalspc}
\end{equation}
where $Y_{i}$ is the cumulative fission yield and $f_{ij}$ is the branching ratio to the daughter level with energy $E_{e}$ and $S_{ij}(E_{\bar{\nu}})$ is the antineutrino spectra for a single transition with endpoint energy $E_{\bar{\nu}}=E_{0}-E_{e}$. The beta-decay spectrum $S_{ij}(E_{\bar{\nu}})$ for a single transition in nucleus (Z,A) with end-point energy $E_{0}$ is
 \begin{equation}
 S(E_{e},Z,A)=S_{0}(E_{e})F(E_{e},Z,A)C(E_{e})[1+\delta(E_{e},Z,A)],
\label{singlespc}
\end{equation}
where $S_{0}(E_{e})=G_{F}^{2}p_{e}E_{e}(E_{0}-E_{e})/2\pi^{3}$, $E_{e}(p_{e})$ is
the electron total energy (momentum), $F(E_{e},Z,A)$ is the Fermi function needed to account for the Coulomb interaction of the outgoing electron with the charge of the daughter nucleus, and $C(E_{e})$ is a shape factor for forbidden transitions due to additional lecton momentum terms. For allowed transitions $C(E_{e})$ = 1. The term
$\delta(E_{e},Z,A)$ represents fractional corrections to the spectrum. The primary corrections to beta decay are radiative $\delta_{rad}$, finite size $\delta_{FS}$, and weak magnetism $\delta_{WM}$.

\begin{figure}
\includegraphics[width=9cm, height=7cm]{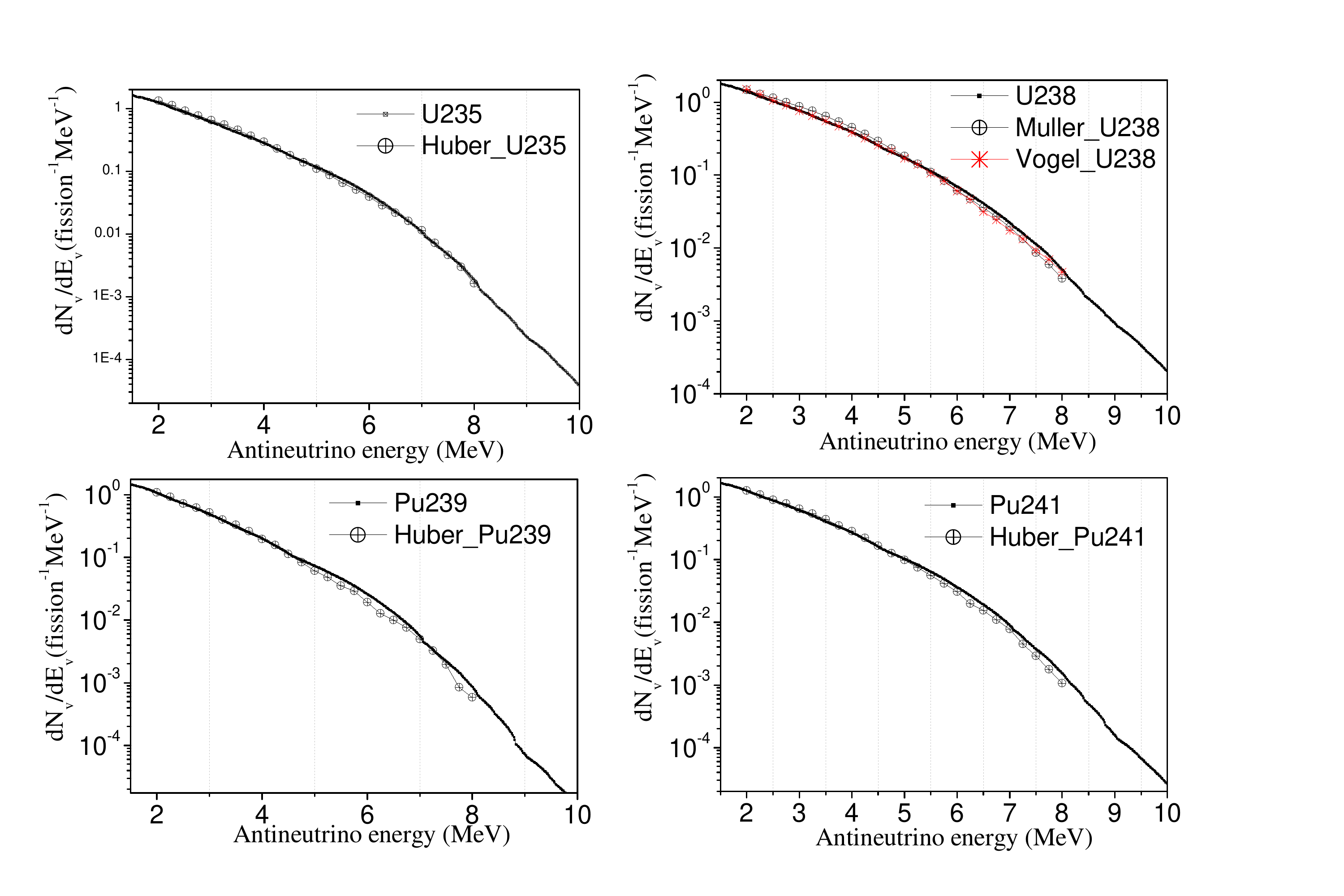}
\caption{\label{u5891spccomp} Calculated antineutrino spectra following the thermal fission of $^{235}$U, $^{239}$Pu, and $^{241}$Pu and the fast fission of $^{238}$U compared with Huber\cite{a11} model, Muller\cite{a7} model and Vogel\cite{avogel} model. }
\end{figure}

 In the present study, we calculated antineutrino spectra following the thermal fission of $^{235}$U, $^{239}$Pu, and $^{241}$Pu and the fast fission of $^{238}$U with the updated database and ENDF/B-VII.1 fission yield\cite{endfyield}.  Fig.\ref{u5891spccomp} shows the spectra comparison with Huber\cite{a11} model, Muller\cite{a7} model and Vogel\cite{avogel} model. As can be seen in Fig.\ref{u5891spccomp}, for $^{239}$Pu, and $^{241}$Pu, the spectra difference in the 5--7 MeV between the calculated spectra and Huber model is obvious.

\begin{figure}
\includegraphics[width=9cm, height=7cm]{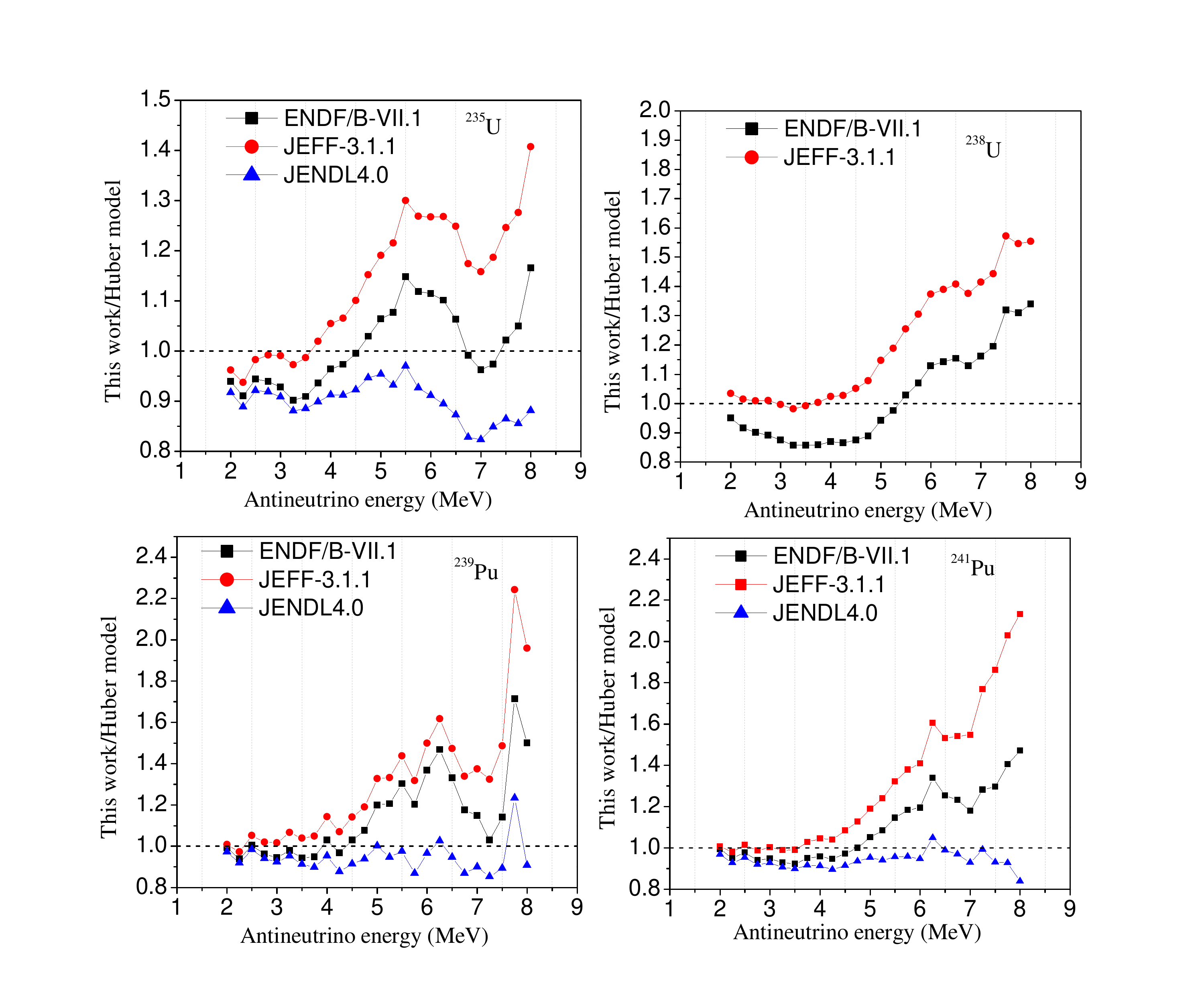}
\caption{\label{Diffdatabase}Spectrum ratio with different fission yield library. In all the cases, the spectrum were calculated from 0 to 10 MeV and the spectrum in the energy region from 2 to 8 MeV were used to compare with Huber-Mueller spectra. The spectra are not normalized to energy region from 2 to 8 MeV.}

\end{figure}

\begin{figure}
\includegraphics[width=9cm, height=7cm]{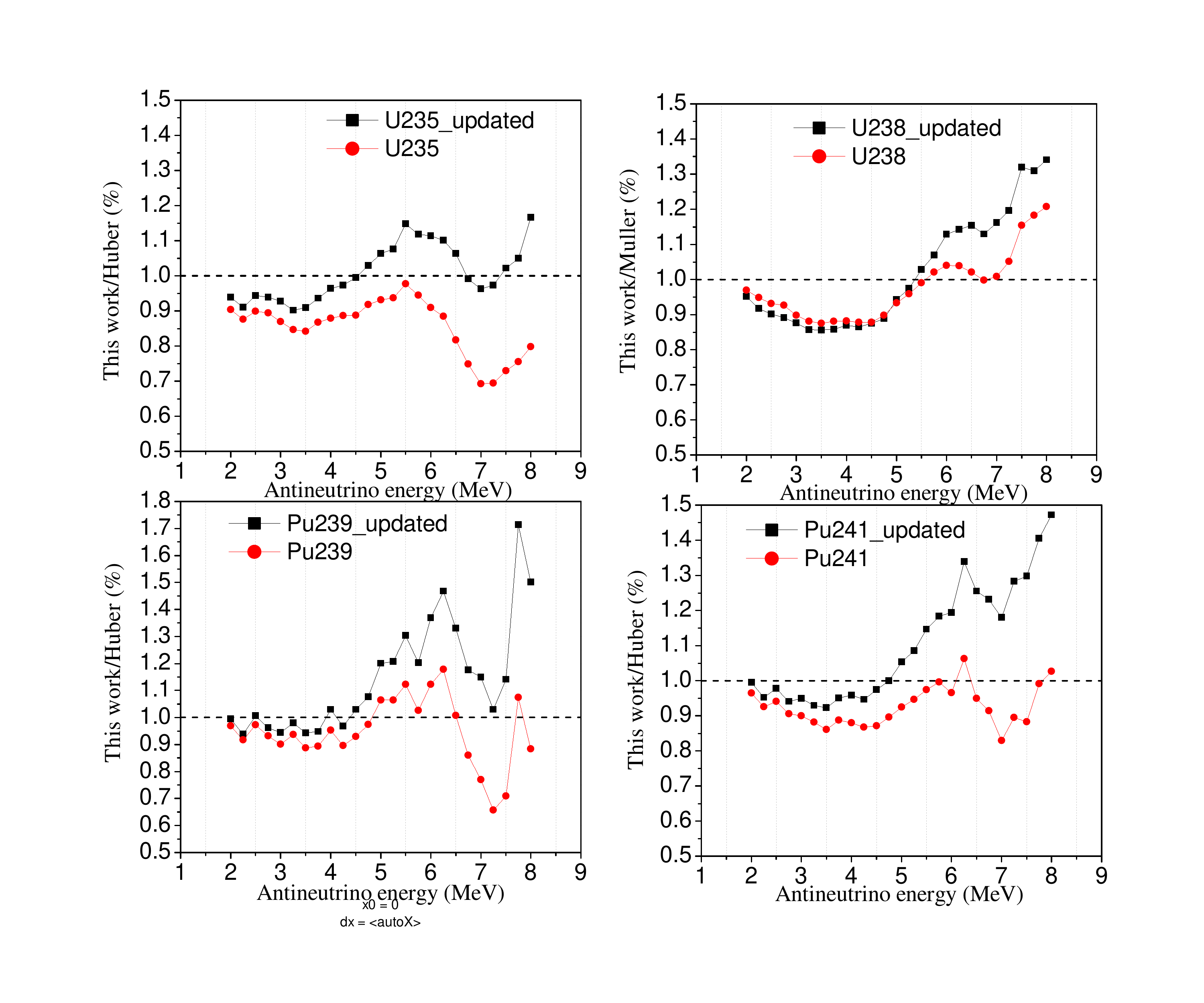}
\caption{\label{u5891ratio2}Comparing the spectra ratio of this study to Huber-Muller model with the updated nuclear database and original JENDL3.1.1. In all the cases, the fission yield data library of ENDF/B-VII.1\cite{endfyield} were used.}
\end{figure}

\section{Spectrum anomaly analysis}
There are four isotopes which are regarded as very important for the reactor antineutrino experiment, such as $^{235}$U, $^{238}$U, $^{239}$Pu and $^{241}$Pu, because the total fission fraction of those four isotopes are more than 99\%. Possible origins of the bump was analyzed and possible explanations were given in Ref.\cite{hayes2015}. They found that the ENDF/B-VII.1 database predicts the antineutrino bump arises from analogous bump in the aggregate fission beta spectra. In contrast, the JEFF-3.1.1 database does not predict a bump. The spectrum ratio with different accumulated fission yield library were shown in Fig.\ref{Diffdatabase}. As shown in the Fig.\ref{Diffdatabase}, the large difference between the two
different fission yield predictions for the bump arises entirely from a difference in the evaluated fission-fragment yields.  For the $^{235}$U, $^{239}$Pu and $^{241}$Pu, the spectra ratio of using ENDF/B-VII.1 is always located between that of using JEFF3.1.1 and that of using JENDL4.0, and the spectra ratio of using JENDL4.0 is relatively flat. The spectra ratio using ENDF/B-VII.1 and JEFF3.1.1 in the high energy region are trend to be higher. For the $^{238}$U, the spectra of Muller was used to be compared. Because there are no $^{238}$U fission yield data in the JENDL4.0, the spectra ratio of this study to Muller model using JENDL4.0 were not given. The difference between this study and Muller's model arises from the different nuclear database.
Fig.\ref{u5891ratio2} show the spectra ratio of this study to Huber-Muller model with the updated nuclear database and original JENDL3.1.1. As can be seen in Fig.\ref{u5891ratio2}, the spectra ratio become large arise from more isotopes having been taken into account in the calculation.

\begin{figure}
\includegraphics[width=8cm, height=6cm]{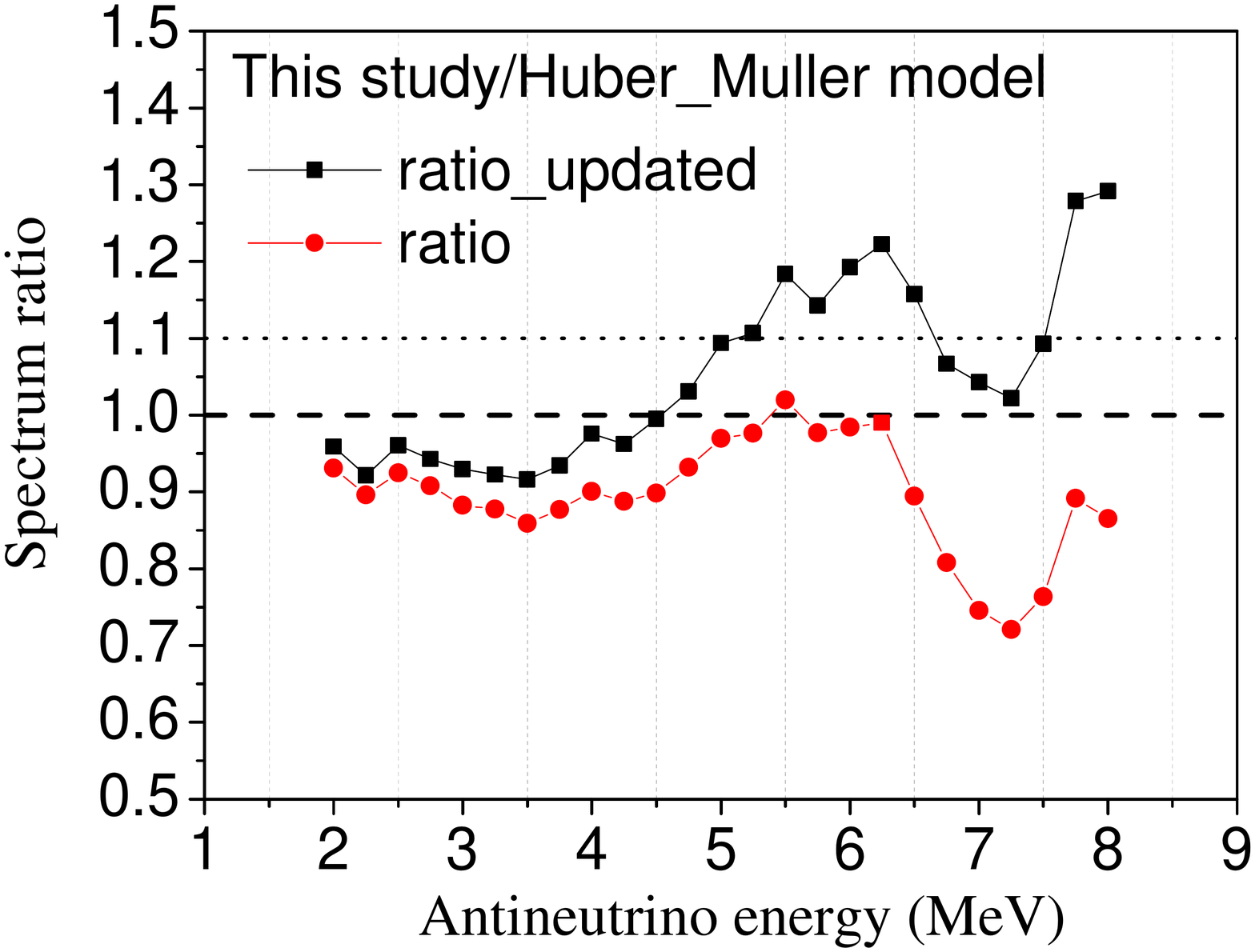}
\caption{\label{TohuberMModel} The total antineutrino spectrum ratio of this work to Huber-Muller model.}
\end{figure}

Fig.\ref{TohuberMModel} shows the updated and original database predictions for the shape of the antineutrino spectra for Daya Bay \cite{spc1Dayabay} relative to the Huber-Mueller model\cite{a7,a11}. In this combination, the fission fraction of $^{235}$U: $^{238}$U : $^{239}$Pu: $^{241}$Pu is 0.586: 0.076: 0.288 :0.05. As can be seen in Fig.\ref{TohuberMModel}, the bump is obvious in the absolute ratio of the ENDF/B-VII.1 aggregate beta spectra to that of Huber-Muller's model. If the original database applied, the absolute ratio can not reach one in most energy region. However, if the updated database applied, the absolute ratio in the low energy region is less than one, but in the high energy region, it is higher than one. The bump was also found in the absolute ratio which has similar structure with the Daya Bay experiment\cite{spc1Dayabay}. Similar spectral structure were also found in ref \cite{dwyer}\cite{hayes}. The original database predictions were almost the same as ref \cite{dwyer}\cite{hayes}, however, after update the database, the bump become more obvious in the energy region from 5--7 MeV.

\section{Important isotope contribute to the bump}
 Several of beta-decay transitions that dominate in the bump region 5--7 MeV have been given in ref \cite{dwyer}\cite{hayes}. In this region the spectral shape is dominated by thirteen prominent decay branches which contribute 56\% of the calculated rate. While the remaining 44\% is composed of more than 1100 decay branches.
 All thirteen branches are transitions between the ground states of the initial and final isotopes, and all are first forbidden nonunique decays. Most prominent beta decay branches with $E_{\bar{\nu}}=5-7$MeV is shown in table \ref{table1}. Beside the eight prominent decay branches given in ref \cite{dwyer}, we found that $^{98,99}$Y, $^{90,94}$Rb and $^{138}$I also have important to the bump, and which have no given in ref \cite{dwyer}.

\begin{table}[b]
\caption{\label{table1}%
Most prominent beta decay branches with $E_{\bar{\nu}}=5-7 MeV$. The table presents the decay parent, end-point energy, half-life, and decay $ft$ value. The rate each branch contributes to the total between $5-7 MeV$ is $N$, accounting for the inverse beta decay cross section.
}
\begin{ruledtabular}
\begin{tabular}{lcdrcc}
\textrm{Isotope}&
\textrm{Q(MeV)}&
$t(s)$ &  Log(ft) &  N(\%) & Ref.\cite{dwyer} \\
\colrule
$^{96}$Y & 7.103 & 5.34 & 5.59 & 12.5 & 13.6 \\
$^{92}$Rb & 8.095 & 4.48 & 5.75& 8.11 & 7.40 \\
$^{142}$Cs &7.308 & 1.68 & 5.59 & 4.89 &5 \\
$^{94}$Rb& 10.283& 2.702 & 7.14 & 4.04&	- \\
$^{99}$Y & 6.969& 1.477 & $>$5.9 & 3.79& - \\
$^{97}$Y & 6.821& 3.75 & 5.7 & 3.42& 3.80 \\
$^{90}$Rb& 6.584& 158 &7.35& 3.26 &- \\
$^{98}$Y & 8.824& 0.548& 5.76& 2.91 &- \\
$^{138}$I& 7.82& 6.46& 8.83& 2.90& - \\
$^{93}$Rb &7.462& 5.84& 6.14& 2.83& 3.70\\
$^{140}$Cs& 6.22& 63.7& 7.05& 2.70 &2.70 \\
$^{100}$Nb& 6.386& 1.5& 5.1& 2.58&	3.00 \\
$^{95}$Sr & 6.089& 23.9& 6.16 &2.47 &2.60 \\
\end{tabular}
\end{ruledtabular}
\end{table}

\section{Average antineutrino spectrum}
\begin{figure}
\includegraphics[width=8cm, height=6cm]{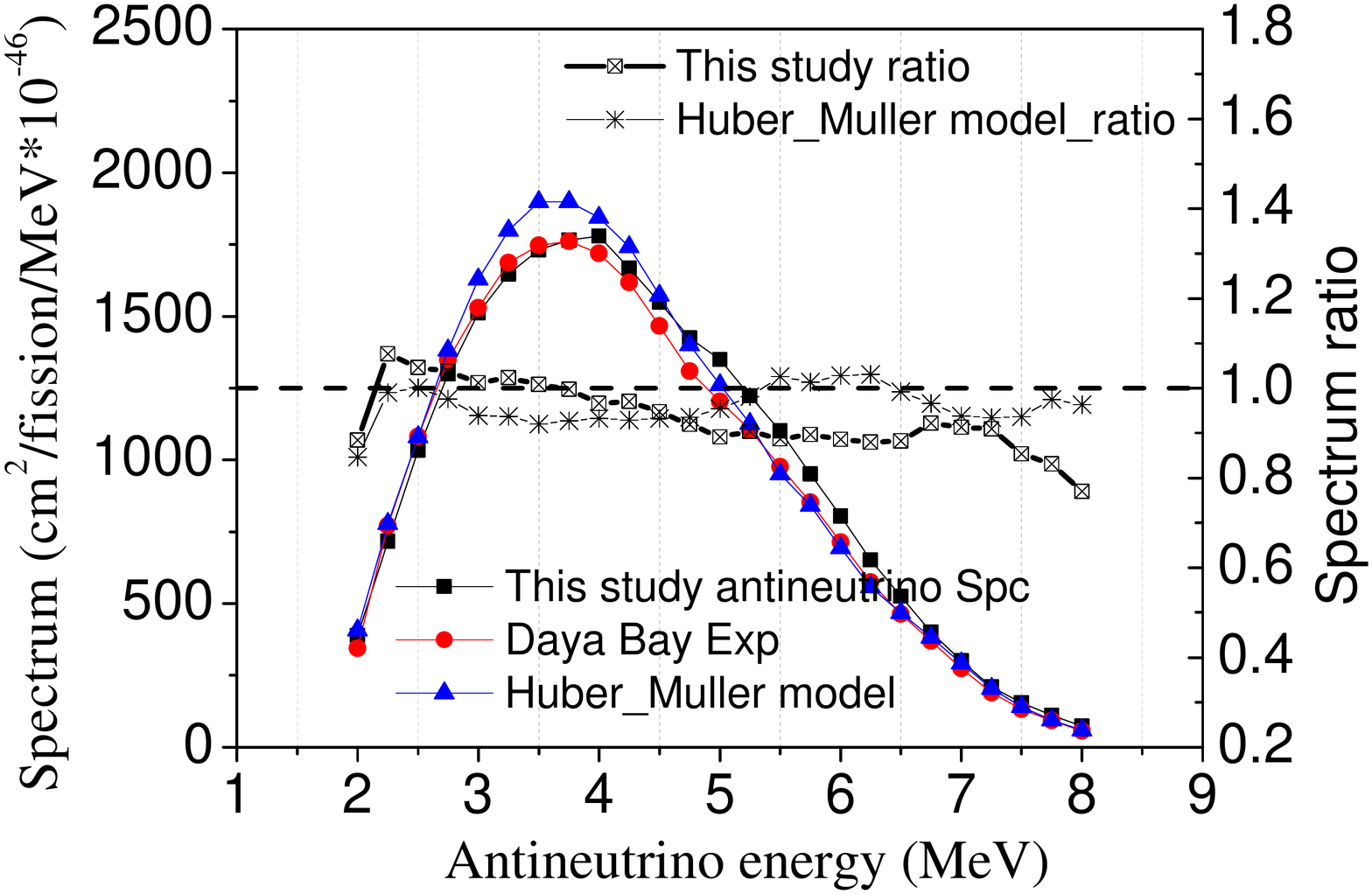}
\caption{\label{spectrum_ratio}(color online). Average spectrum of this study, Daya Bay experiment and Huber-Muller model are shown in black, red and blue line. Spectra ratio of $Ab ~ initio$ nuclear calculation cumulative antineutrino spectra to the Huber-Muller model and Daya Bay experiment.}
\end{figure}

Applying Daya Bay experiment fission fraction of $^{235}$U: $^{238}$U : $^{239}$Pu: $^{241}$Pu is 0.586: 0.076: 0.288 :0.05, the average antineutrino spectrum of this study and Huber-Muller model were calculated, and the Daya Bay experiment spectrum were given in ref \cite{spc1Dayabay}. The spectra were shown in Fig.\ref{spectrum_ratio}. As shown in Fig.\ref{spectrum_ratio}, the spectrum ratio of Daya Bay experiment to this study and Huber-Muller model have the trend on the contrary. The spectrum ratio of Daya Bay experiment to this study is relative higher in the low energy region, but relative smaller in the higher energy region. On the other hand, The spectrum ratio of Daya Bay experiment to Huber-Muller model is lower in the low energy region, but higher in the high energy region, and then we see the bump. In the $Ab ~initio$ method, we assume that each nuclides are in equilibrium, and the decay rate of each nuclides can be replaced by the cumulative yield, but in the reactor, they are different because many kind type of reaction have the possible to be happen, which may be the reason for the bump.

\begin{figure}
\includegraphics[width=8cm, height=9cm]{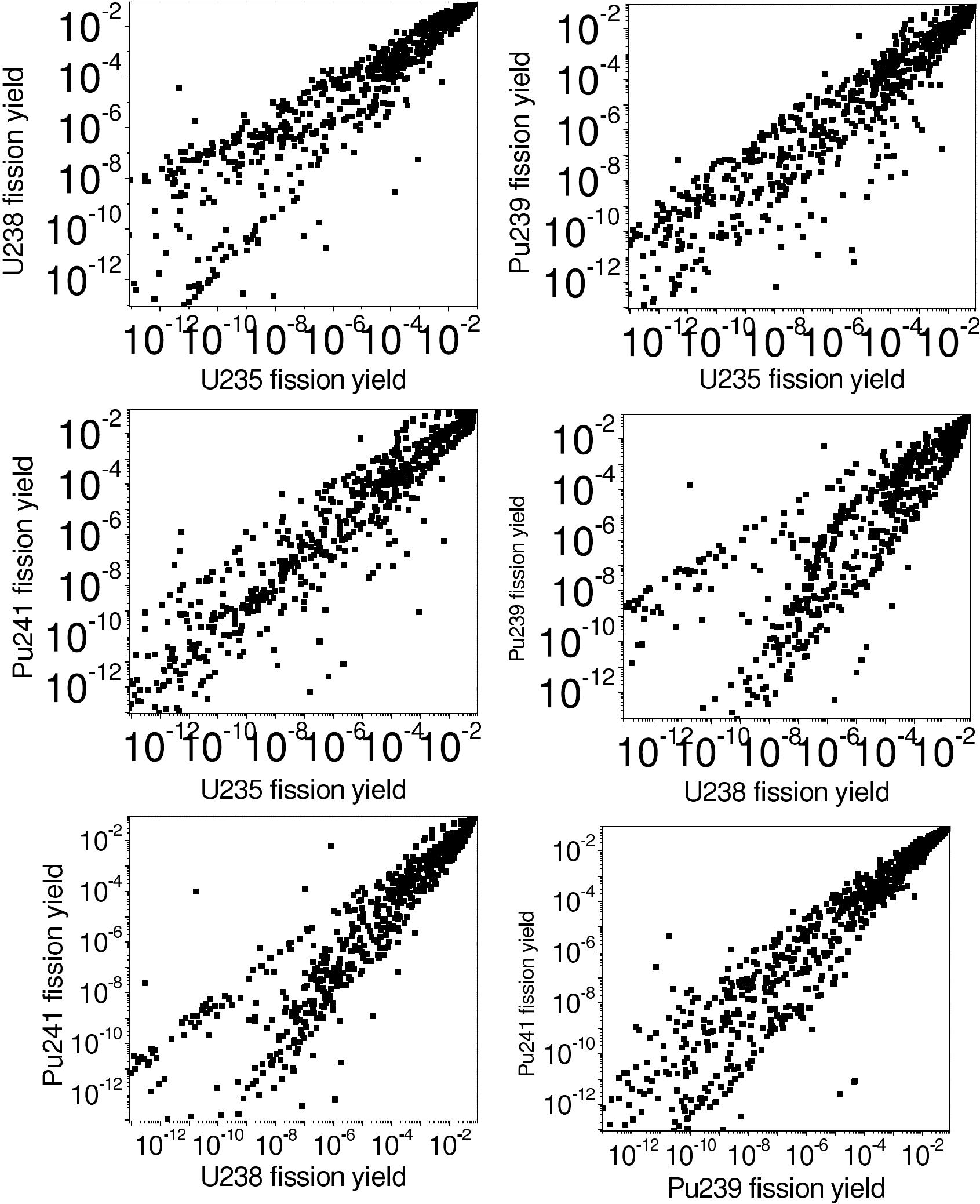}
\caption{\label{correlation} Fission yield of isotopes $^{235}$U, $^{238}$U, $^{239}$Pu and $^{241}$Pu.}
\end{figure}
\begin{figure}
\includegraphics[width=8cm, height=6cm]{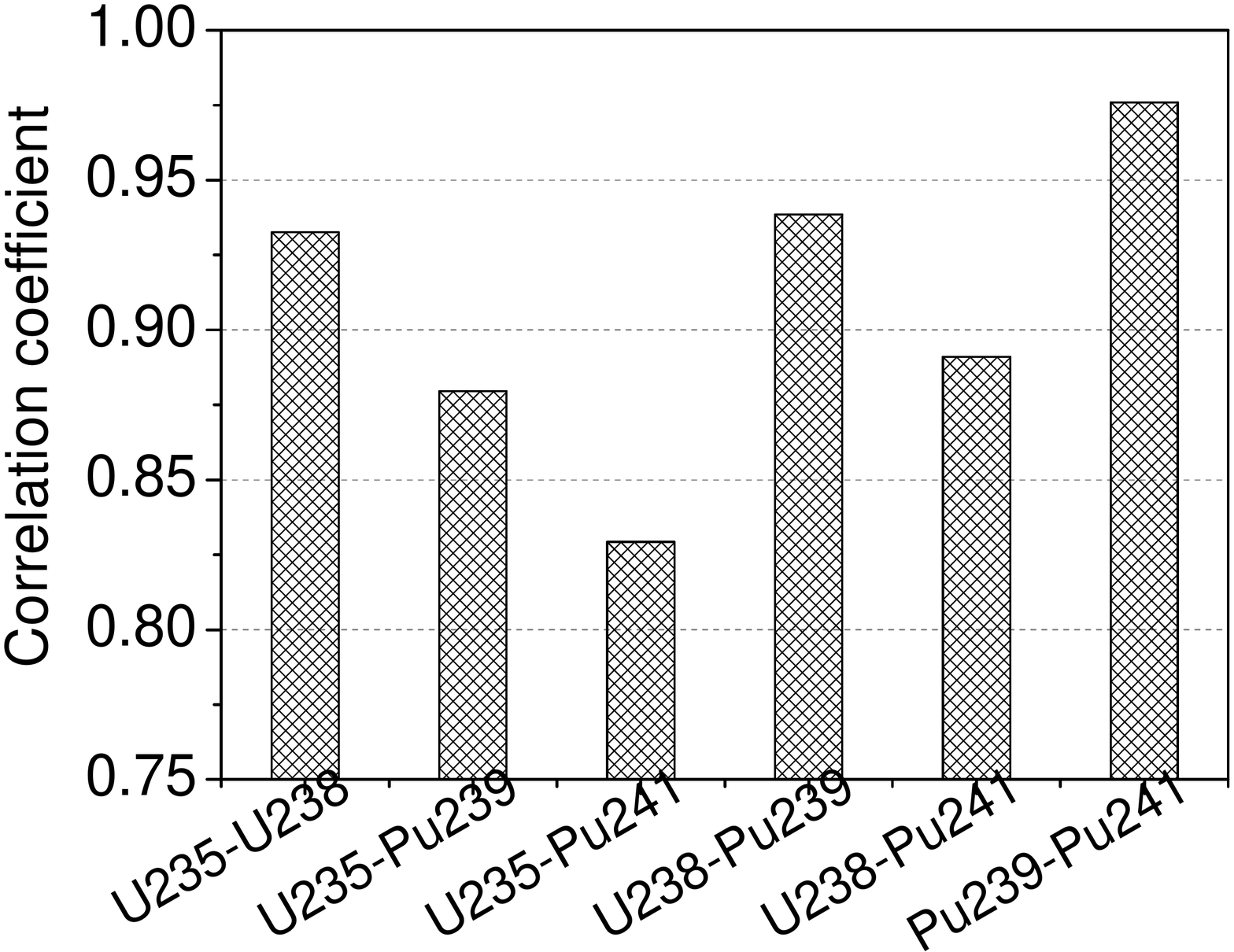}
\caption{\label{correlationiso} Fission yield correlation coefficient between four isotopes .}
\end{figure}

The Daya Bay experiment has observed correlations between reactor core fuel evolution and changes in the reactor antineutrino flux and energy spectrum\cite{fuelEvoDayabay}. The measured evolution in total inverse beta decays (IBD) yield disagrees with recent predictions at 3.1$\sigma$. This discrepancy indicates that an overall deficit in the measured flux with respect to predictions does not results from equal fractional deficits from the primary fission isotopes $^{235}$U, $^{238}$U, $^{239}$Pu, $^{241}$Pu. The Daya Bay results shows that $^{235}$U may be the primary contribution to the antineutrino anomaly because of a 7.8\% discrepancy between the observed and predicted yields. For the RENO experiment, the measured IBD yield per $^{235}$U fission shows the largest deficit relative to a reactor model prediction, and they report a hint of correction between the 5 MeV bump in the IBD spectrum and the reactor fuel isotope fraction of $^{235}$U\cite{fuelEvoRENO}.  Fig.\ref{correlation} show the fission yield correlation between the isotopes $^{235}$U, $^{238}$U, $^{239}$Pu and $^{241}$Pu. As can be seen in the Fig.\ref{correlation}, there have strong correlation between the four isotopes, and this can be seen from the Fig.\ref{correlationiso}. Most correlation coefficient between the four isotopes are more than 0.85, and the largest correlation coefficient is taken place between $^{239}$Pu and $^{241}$Pu, and which is 0.96. Although, a hint of correction between the 5 MeV bump in the IBD spectrum and the reactor fuel isotope fraction of $^{235}$U, the other isotopes may also have correlation with the bump because of large fission yield correlation coefficient.

\section{Conclusion and discussion}
Although, the database analyses of the antineutrino spectra provide any insight into the reactor neutrino anomaly, and the bump was found with the updated database, the database prediction uncertainties are too large to draw any conclusion. The spectrum ratio of Daya Bay experiment to the ENDF/B-VII.1 prediction are close to one, and no bump appeared. More important isotopes are fond to contribute to the bump, and very isotope contribution is more than 2.4\%. According to the fission yield correlation between the isotopes $^{235}$U, $^{238}$U, $^{239}$Pu and $^{241}$Pu, strong correlation are found between those isotopes, and most correlation coefficient are more than 0.85. Further bump or anomaly analyses should be taken into account those correlations.

Both the anomaly and the bump problems may due to problem with the original aggregate beta-spectra measurement\cite{hayes2015} at the ILL, or the conversion process from beta spectra to antineutrino. It is difficult to answer those questions with current theoretical frameworks or from existing data. Consequently a better idea is to do the isotope unfolding from the Daya Bay experiment or carry out an experiment at CSNS\cite{CSNS} to measurement the electron spectrum more precisely again.




\end{document}